\documentclass[12pt,preprint]{aastex}

\usepackage{psfig}
\usepackage{longtable}
\def\hb{\ifmmode {\rm H}\beta \else H$\beta$\fi}

\def \L5100{$L_{5100}$}

\def \ir20551{IRAS~20551$-$4250}
\def \mum{~$\mu$m}
\def \h2{H$_2$}

\slugcomment{}

\shorttitle{IRAS~20551$-$4250 circumnuclear environment} 
\shortauthors{E. Sani \& E. Nardini}

\begin{document}

\title{The circumnuclear environment of IRAS~20551$-$4250:\\
a case study of AGN/Starburst connection for \textit{JWST}}

\author{E. Sani,\altaffilmark{1} E. Nardini \altaffilmark{2}}

\email{sani@arcetri.astro.it}
\email{enardini@cfa.harvard.edu}

\altaffiltext{1}{INAF - Osservatorio Astrofisico di Arcetri, Largo
E. Fermi 5, I-50125 Firenze, Italy}
\altaffiltext{2}{Harvard-Smithsonian Center for Astrophysics, 60
Garden Street, Cambridge, MA 02138}

\begin{abstract}
We present a general review of the current knowledge of \ir20551 and its circumnuclear environment. 
This Ultraluminous Infrared Galaxy is one of the most puzzling sources of its class in the nearby 
Universe: the near-IR spectrum is typical of a galaxy experiencing a very intense starburst, 
but a highly obscured active nucleus is identified beyond $\sim$5~$\mu$m and possibly dominates 
the mid-IR energy output of the system. At longer wavelengths star formation is again the main 
driver of the global spectral shape and features. We interpret all the available IR diagnostics 
in the framework of simultaneous black hole growth and star formation, and discuss the key 
properties that make this source an ideal laboratory for the forthcoming {\it James Webb 
Space Telescope}. 
\end{abstract}

\keywords{galaxies: active - galaxies: starburst - infrared: galaxies - galaxies: individual (IRAS~20551$-$4250)}


\section{Introduction}
Two main physical processes characterize the nuclear regions of active galaxies: intense star formation 
at rates of $\sim$10$^2$--10$^3$ M$_\odot$~yr$^{-1}$ (starburst, SB) and accretion on to a supermassive 
black hole (active galactic nucleus, AGN). The issue of SB and AGN connection in both local and distant 
galaxies is critical for a proper understanding of galaxy formation and evolution, of star formation 
history and metal enrichment of the Universe, and of the origin of the extragalactic background at low 
and high energies. There is indeed increasing evidence of a strong link between the starburst and AGN 
mechanisms in active systems. The empirical correlation between the mass of black holes (BHs) located at 
the centre of nearby galaxies (both active and passive/quiescent) and the mass of their spheroids (see Sani 
et al.~2011 [1] and references therein) suggests that the formation of bulges and the growth of the central 
BHs are tightly connected. Also the presence of circumnuclear star formation in a substantial fraction of 
local AGN (Genzel et al.~1998 [2], Cid Fernandes et al.~2004 [3], Schweitzer et al.~2006 [4], Sani et 
al.~2010 [5]) hints at the relation between the two phenomena. The overall conclusion of these studies 
is that in 30--50\% of the cases the accreting supermassive BHs are associated with young (i.e. of age 
less than a few $\times$100 Myr) star-forming regions, with clear evidence of an enhanced star 
formation rate (reaching up to starburst intensities) in most AGN. However, this does not necessarily 
imply any causal connection between the two physical processes. It could be simply the natural 
consequence of massive gas fuelling into the nuclear regions, due to either interactions/mergers or 
secular evolution such as bar-driven inflows. Both star formation and nuclear accretion, in fact, are 
triggered and subsequently fed by this gas reservoir.\\
In the local Universe, the optimal targets to study the AGN/SB interplay are the so-called Ultraluminous 
Infrared Galaxies (ULIRGs; Sanders \& Mirabel 1996 [6]). These sources are the result of major mergers, during 
which the redistribution of the gaseous component drives vigorous starburst events and obscured nuclear 
accretion. It is now well established that ULIRGs are usually powered by a combination of both processes, 
giving rise to their huge luminosities ($L_{\rm{bol}} \sim L_{\rm{IR}} > 10^{12} L_\odot$). However, since 
the primary radiation field is reprocessed by dust, the identification of the dominant energy supply is often 
unclear. The simultaneous presence of star formation and AGN signatures in the mid-IR makes this a really 
favourable band to disentangle the AGN and SB components and explore their environment. 
In particular, (\emph{i}) the available spectra of \textit{bona fide} starburst-dominated and, respectively, 
unobscured AGN-dominated sources are widely different, and show little dispersion within the separate classes 
(Risaliti et al. 2006 [7], Brandl et al. 2006 [8]; Netzer et al. 2007 [9], Nardini et al. 2008 [10]). 
This allowed us to reproduce the AGN/SB contributions with fixed templates, especially 
over the the 3--8\mum\ spectral interval. 
(\emph{ii}) For a given bolometric luminosity, the mid-IR AGN emission is higher than that of a starburst 
by a factor that rapidly declines with wavelength, ranging from $\sim$100 at 3--4~$\mu$m [7] to $\sim$25 at 
5--8~$\mu$m (Nardini et al.~2009 [11]). Such a large difference is due to the key contribution of the hot dust 
layers directly exposed to the AGN radiation field. Together with the relatively low dust extinction at these 
wavelengths, this allows the detection of an AGN even when it is heavily obscured and/or its total luminosity is 
small compared with the SB counterpart. Based on the above points, we successfully fitted the observed ULIRG 
spectra with a two-component analytical model, with only two free parameters: the relative AGN/SB contribution 
and the optical depth of the screen-like obscuration (if any) affecting the compact AGN component.\\ 
To understand whether the link between star formation and nuclear activity is a matter of \textit{nature} (i.e. 
feedback processes) or \textit{nurture} (i.e. host environments), here we investigate the circumnuclear 
structure of \ir20551, an ideal laboratory thanks to its unique physical properties (in terms of both relative 
AGN/SB contribution and AGN obscuration), and to the fairly large multiwavelength dataset available. 
The paper is organized as follows: in Section 2 we review the present knowledge of the mid-IR properties of 
\ir20551. The dust extinction law and gas column density are dealt with in Section 3. A possible general 
picture and the feasibility of future observations with {\it James Webb Space Telescope} (\textit{JWST}) are 
discussed in Section~4. In Section~5 we summarize our findings and draw the conclusions. Throughout this work we 
adopt a standard cosmology ($H_0=70$~km/s/Mpc, $\Omega_m=0.3$, $\Omega_\lambda=0.7$).
\begin{figure}[!h]
  \centering
   \includegraphics[angle=0,width=16cm]{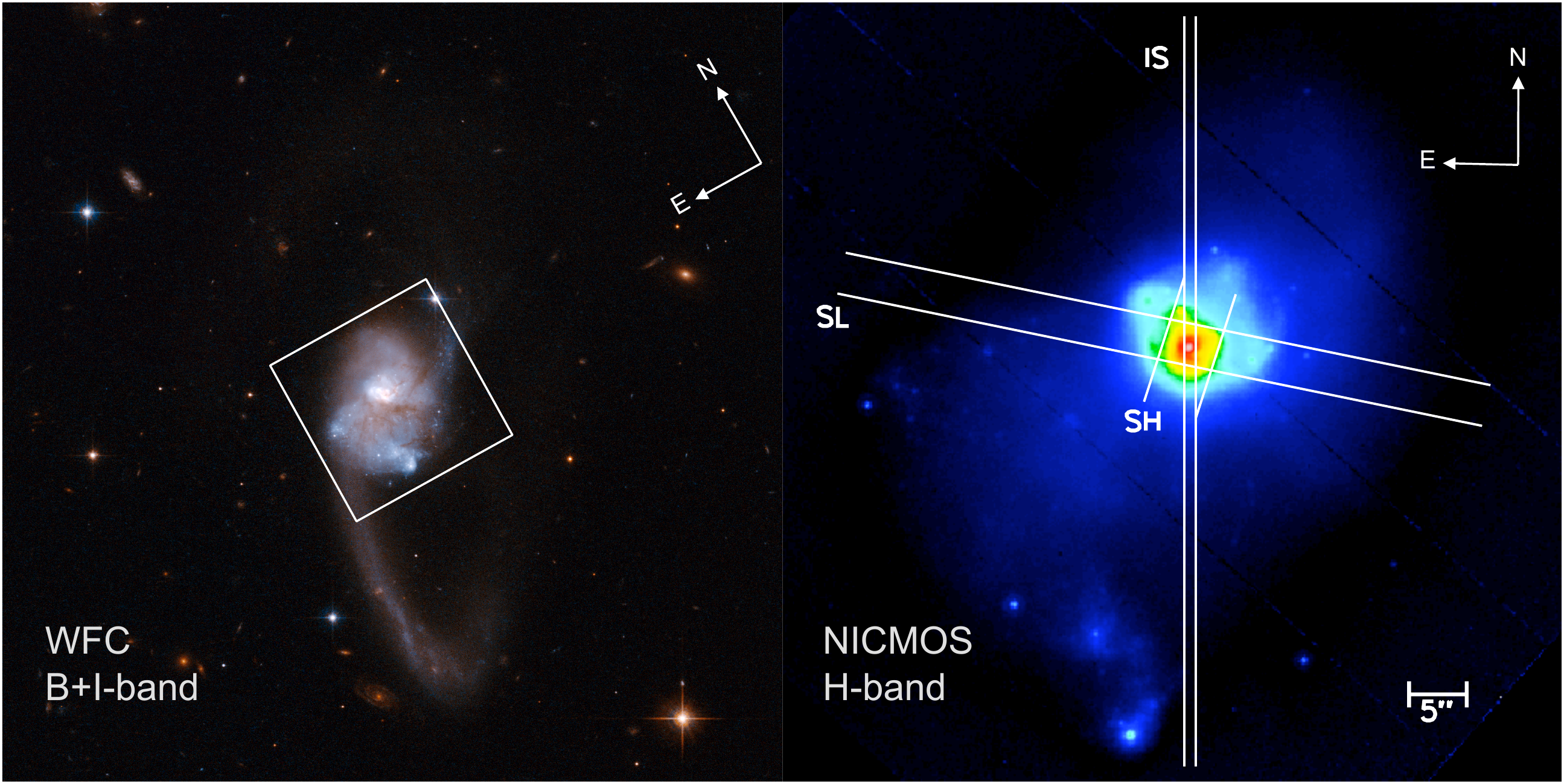}
    \caption{{\footnotesize \textit{HST} images of \ir20551. Left: composite \textit{B}+\textit{I} band image 
obtained with the WFC F435W and F814W filters 
    (PID 10592, PI A. Evans). 
    The white square identifies the nuclear ($30\arcsec \times 30\arcsec$) region. Right: NICMOS F160W image (PID 11235, PI J. Surace) 
    of the nuclear region. 
    The \textit{Spitzer}/IRS short-wavelength/low-resolution (SL, $3.6\arcsec$ width) and short-wavelength/high-resolution 
    (SH, $4.7\arcsec$ width) slits are shown together with the 
    \textit{VLT}/ISAAC slit (IS, $1\arcsec$ width in the \textit{L}-band). Given a spatial scale of $\rm 950~pc/\arcsec$, 
    the SL, SH and IS slits cover regions of 
    3.4~kpc, 4.5~kpc and 950~pc respectively.}}
\label{fig0}
\end{figure}
\section{\ir20551: general properties}
\ir20551\ is a nearby ($z=0.043$) ULIRG lying in the luminosity range of IR quasars, with 
L$_{IR}=4.3\times10^{45}$~erg/s. It mostly lacks of targeted studies, none the less 
in literature there are several related measurements among the statistical analyses of the 
local ULIRG population. \ir20551\ is a merging system in a fairly advanced state (Fig~\ref{fig0}, 
left panel), characterized by a single nucleus with a prominent tidal tail and a slightly disturbed 
core, likely caused by a minor merger or strong secular evolution effects. 
From the high resolution near-IR data Haan et al. (2011) [12] ascribe the large ratio of nuclear excess to 
bulge luminosity (see also Fig.~\ref{fig0}, right panel) to the possible presence of AGN with BH mass 
$\sim 4.4\times10^8 \rm M_\odot$. 
The spectral classification changes significantly with the observed waveband. It is optically 
classified as a H~\textsc{ii} region (Kewley et al. [13]), while in the mid-IR it resembles a SB galaxy [2]. 
However, diagnostic methods exclusively based on emission lines, as the ones mentioned above, suffer 
from limited extensibility to faint sources and fail in identifying the heavily absorbed AGN detected 
in the hard X-rays. Indeed, the hard X-rays emission of \ir20551\ is clearly dominated by an obscured AGN, 
with luminosity $L_{2-10\;keV}\sim~7.0\times10^{42}$ erg~$s^{-1}$ and column density 
N$_H\sim8\times10^{23}$~cm$^{-2}$ (Franceschini et al. 2003 [14]). 
According to all these pieces of observational evidence, the relative AGN contribution to the bolometric 
luminosity is uncertain, but probably highly significant, while the circumnuclear environment is still poorly 
characterized. 
The first quantitative determination of the AGN contribution to the mid-IR emission of \ir20551\ was obtained  
by Farrah et al. (2007, [15]) thanks to a series of effective diagnostics based on fine-structure lines. 
Their analysis of \textit{Spitzer}/IRS high-resolution spectra suggests a moderate AGN contribution, even though 
a peculiar geometry and/or extreme optical depth are responsible for the lack of typical AGN tracers (e.g. [Ne~\textsc{v}], [O~\textsc{iv}]). 
 
\subsection{\textit{L}- and \textit{M}-band spectroscopy}
Risaliti, Imanishi \& Sani [16] obtained \textit{L}-band observations of ULIRGs with 8-m class telescopes (\textit{VLT} and 
\textit{Subaru}). 
The resulting high-quality spectra have revealed the great power of \textit{L}-band diagnostics in characterizing AGN and SB components inside ULIRGs. 
The main results of these studies are summarized in the following. 
\textit{(1)} A large ($\sim 110$~nm) equivalent width (EW) of the 3.3 $\mu$m polycyclic aromatic hydrocarbon 
(PAH) emission feature is typical of SB-dominated sources, while the strong radiation field of an AGN, extending 
up to the X-ray domain, partially or completely destroys the PAH carriers. 
\textit{(2)} A strong ($\tau_{3.4}>0.2$) absorption feature at 3.4~$\mu$m due to aliphatic hydrocarbon grains is 
an indicator of an obscured AGN; indeed, such a deep absorption requires the presence of a bright, point-like 
source behind a screen of dusty gas. 
\textit{(3)} A steep continuum ($\Gamma>3$ when describing the flux density as a power law 
$f_\nu \propto \lambda^\Gamma$) hints at the presence of a highly-obscured AGN. Again, a large value of 
$\Gamma$ implies the strong dust reddening of a compact source.\\
The \textit{L}-band spectrum of \ir20551\ shows somewhat puzzling properties [7]: a strong 3.3~$\mu$m emission 
feature (EW $\simeq$ 90 nm) suggests a dominant starburst contribution. 
On the other hand, the steep observed slope ($\Gamma \sim 5$) and the detection of the 3.4-$\mu$m absorption 
feature point to the presence of a significant AGN affecting the continuum emission. 
Sani and co-authors [17] added the \textit{M}-band (4--5~$\mu$m) data to better determine the continuum trend 
and analyse the broad CO absorption band near $4.65~\mu$m. 
\begin{figure}[!h]
  \centering
   \includegraphics[angle=0,width=16cm]{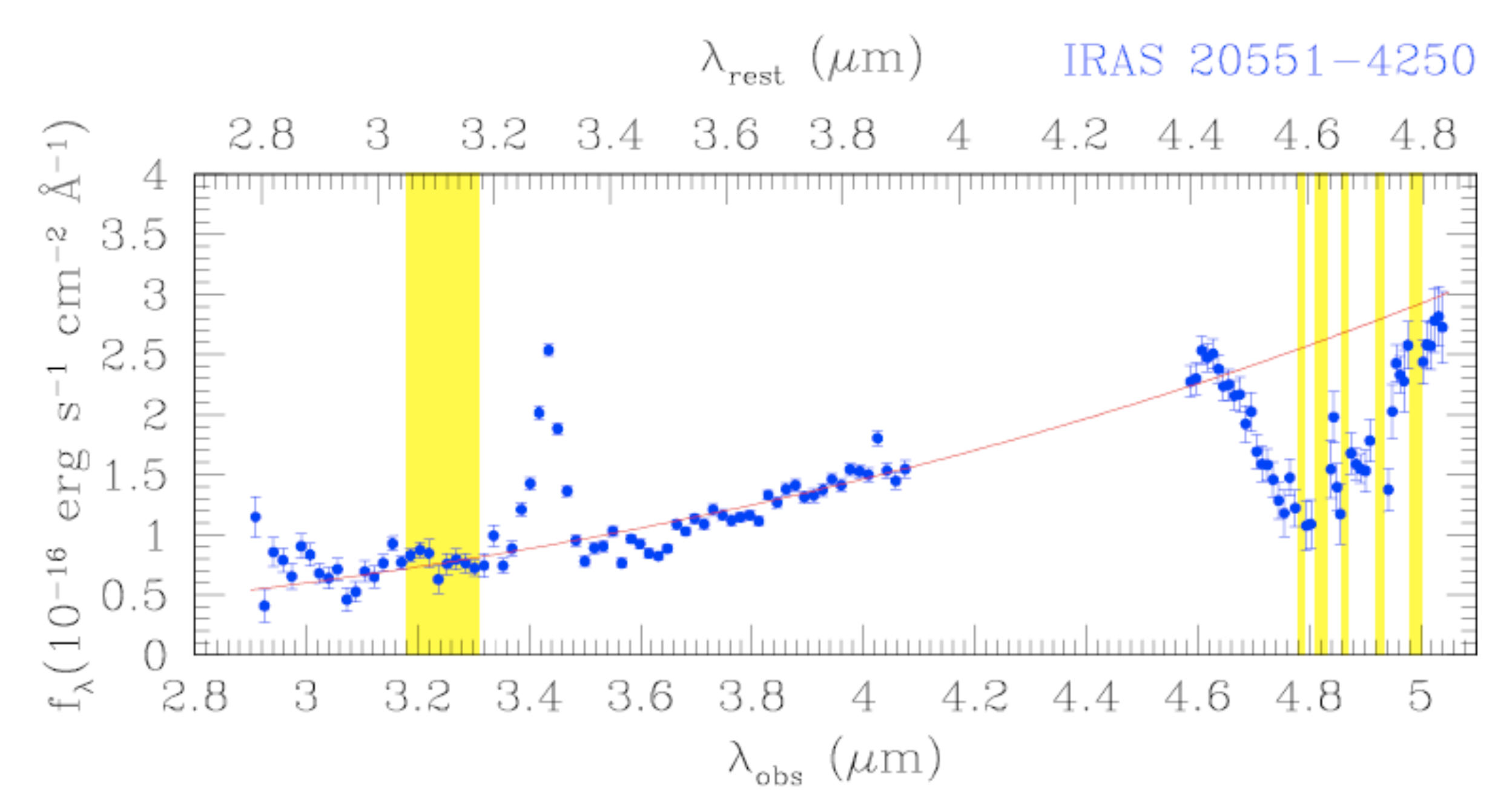}
    \caption{{\footnotesize \textit{L}+\textit{M} band of \ir20551\ [17]. The heavily reddened continuum (red curve) is 
present together with strong CO absorption in the \textit{M}-band. The regions of bad atmospheric transmission are shaded 
in yellow. (Note the units on the vertical axis, $f_\lambda \propto \lambda^{-2} f_\nu$).}}
\label{fig1}
\end{figure}
By combining the \textit{L}- and \textit{M}-band data (as shown in Fig~\ref{fig1}), we estimated a very large 
AGN contribution at 3.5~$\mu$m, exceeding $\sim$90\% once corrected for extinction (see [7] for the analytical 
details). The observed AGN component, however, is heavily obscured and shows extreme dust reddening. The large 
optical depth ($\tau_L > 5$, assuming the extinction law of Draine 1989 [18]) is necessary to reconcile the 
apparently contradictory observational results, i.e. the high equivalent width of the 3.3-$\mu$m PAH feature and 
the steep, intense continuum. The presence of a dust and gas screen absorbing the AGN emission is also revealed 
by the deep absorption profiles due to aliphatic hydrocarbons ($\tau_{3.4}=1.5$) and gaseous CO 
($\tau_{4.6}=2.2$). This step-wise correlation between continuum reddening and absorption features appears to be 
a general property of ULIRGs hosting an obscured AGN [11,17]. Anyhow, this does not hold under a quantitative 
point of view: no tight correlation is found among the values of the optical depth, not even between the two 
absorption features themselves. This suggest a non-uniform dust composition among ULIRGs. The implications on 
the shape of the extinction law are discussed in the following section.
\subsection{\textit{Spitzer}/IRS spectroscopy} 
In a series of papers [10,11] we have shown that the high quality of \textit{Spitzer}-IRS data 
allows a very effective \textit{quantitative} determination of the AGN/SB components around 5--8~\mum; 
this method is much more accurate than those possible in other bands in spite of the lower AGN over SB 
brightness ratio, which rapidly declines with wavelength. 
Summarizing, once applied to large, virtually complete samples of local ULIRGs, the 5--8\mum\ analysis yields 
the main results listed below: 
\textit{(1)} The large variations in the observed spectral shape of ULIRGs can be successfully explained in 
terms of the relative AGN contribution and its degree of obscuration. 
\textit{(2)} Although the larger fraction of ULIRG bolometric energy output is associated with the intense SB 
events, the AGN contribution is non-negligible ($\sim$25--30\%) and increases with both the total IR 
luminosity of the host galaxy and, possibly, with the merger stage (Nardini et al. 2010 [19]). 
\textit{(3)} The apparent lack of continuum reddening and the simultaneous detection of deep absorption 
troughs in some of the most obscured sources (when a step-wise correlation is generally found, as mentioned 
earlier) suggests that the extinction of the AGN component in a ULIRG environment is not universal. Both a 
power-law and a quasi-grey behaviour of the optical depth as a function of wavelength are necessary to account 
for the emission of different objects and seem to be involved among ULIRGs.\\
\begin{figure}[!h]
  \centering
   \includegraphics[angle=0,scale=0.7]{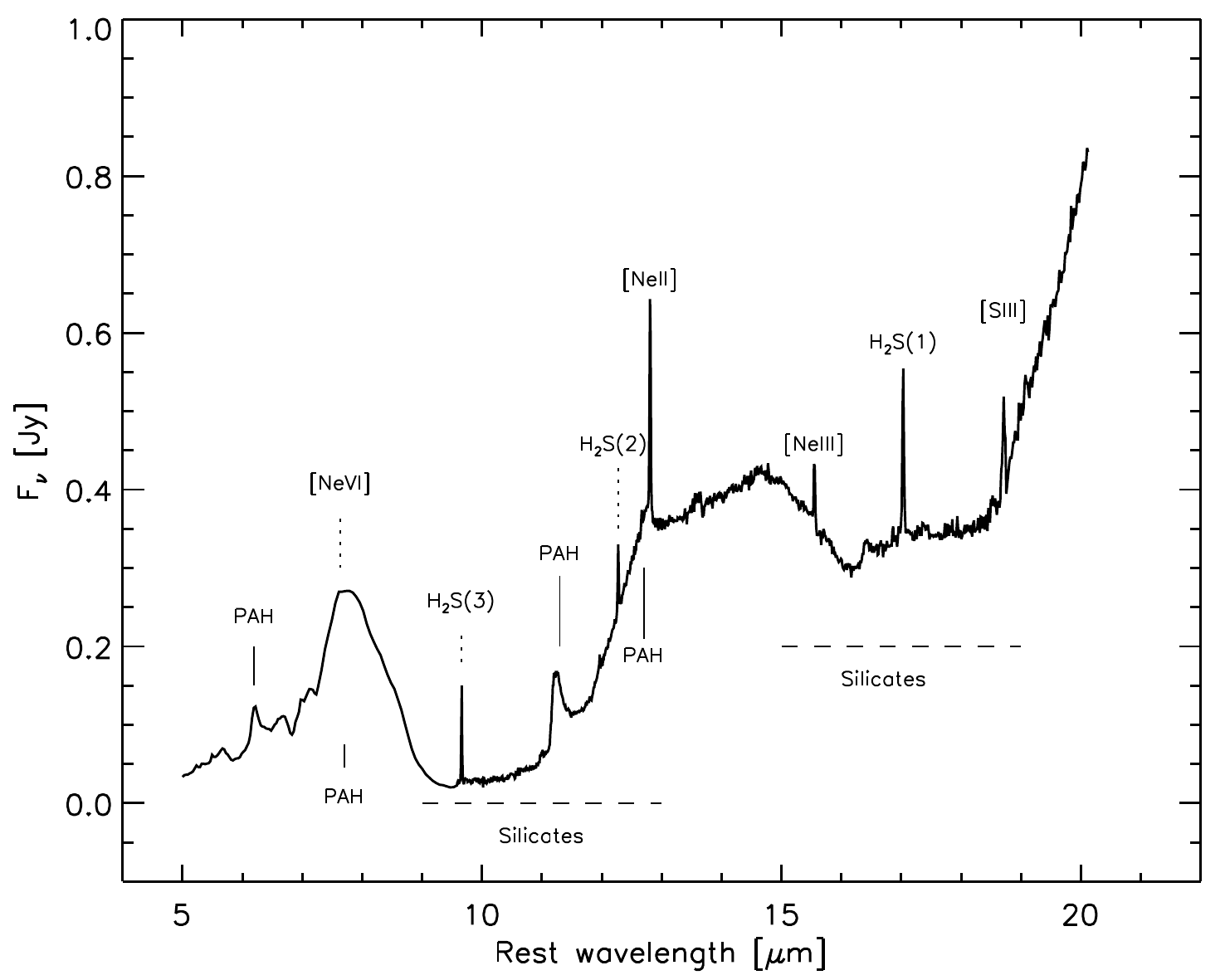}
    \caption{{\footnotesize \textit{Spitzer}/IRS 5--20\mum\ emission. We have already analysed the 
low-resolution data in a previous work [11], while the high-resolution spectrum (above $\sim$10~$\mu$m) has been 
extracted from the same dataset following Schweitzer et al. [4]. The main features are labelled for ease of 
identification.}}
\label{fig2}
\end{figure}
Consistently with the 3--4\mum\ analysis, also the 5-8\mum\ spectrum of \ir20551\ (in Figs.~\ref{fig2} and 
\ref{fig3}) shows remarkable properties: 
the AGN continuum can be hardly determined due to strong absorption around 6 and 6.85~$\mu$m, respectively 
attributed to a mixture of ices and hydrogenated amorphous carbons (HAC). The standard spectral decomposition 
yields again a very bright but strongly reddened AGN, with a mid-IR intrinsic contribution of $\sim$90\% and a 
6-$\mu$m optical depth $\tau_6=1.2$ (following the same extinction law introduced before [18]). 
Although The starburst dominates the bolometric luminosity, the AGN contribution is significant (26$\pm$3\%). 
At longer wavelengths ($\lambda > 8$~$\mu$m), the huge silicate absorption troughs at 9.7 and 18\mum\ 
require the nuclear source to be deeply embedded in a smooth distribution of dust, both geometrically and 
optically thick. Ground-based imaging at 18\mum\ reveals a compact unresolved source ($<120$~pc) with high 
surface brightness and large Si optical depth ($\tau_{18}=0.7$), in agreement with a buried AGN interpretation 
(Imanishi et al. 2011 [20]). It is also worth noting that $\tau_{9.7}$ can be combined with the EW of the 
6.2-$\mu$m PAH feature in a diagnostic diagram that provides not only a direct classification, but also possible 
indications on the evolutionary path of a source, by probing the age of the SB and the geometrical structure of 
the dust (Spoon et al. 2007 [21]). The location of \ir20551\ in such a diagram is typical of an intermediate 
stage between a fully obscured AGN and an unobscured nuclear starburst.\\ 
\begin{table*}
 \begin{center}
 \begin{footnotesize}
 \begin{tabular}{ccccccccc}
 \multicolumn{9}{c}{Mid-IR emission lines}\\
 \hline
  Line                                & H$_2$S(3) & H$_2$S(2) & [Ne II]   & [Ne III]   & H$_2$S(1) & [S III]     & H$_2$S(0) & [S III]\\
  $\lambda_{rest}$(\mum) & 9.662         & 12.275       & 12.814 & 15.555 & 17.030       & 18.713  & 28.219      & 33.481 \\
  E$_{ion}$(eV)                 &  -                & -                & 21.6      &  41.0    &  -               & 23.3       & -               &   23.3  \\
  Flux  ($10^{21}$~W/cm$^{2}$)  &  5.36$\pm$0.15 & 3.06$\pm$0.17 & 13.0$\pm$0.4 & 2.6$\pm$0.3 & 6.9$\pm$0.2 & 
               5.7$\pm$0.6 & 3.8$\pm$0.8 & 8.0$\pm$0.3 \\
  \hline
  \end{tabular}
   \label{tab1}
  \caption{{\footnotesize. Properties of the detected mid-IR emission lines. 
  We measured the fluxes with the {\scshape idl} package {\scshape smart} by means of a Gaussian fitting. 
    We made use the most recent version of the IRS pipelines, thus our estimates are more reliable than those 
    previously published.}}
    \end{footnotesize}
 \end{center}
\end{table*}
As mentioned before, in \ir20551\ also fine-structure lines from highly-ionized atoms are detected, as well as 
H$_2$ pure vibrational transitions (Fig.~\ref{fig2}). 
Our new measurements of the mid-IR line fluxes are listed in Tab.~1.  
Notably, the standard coronal lines produced by the hard AGN photons, such as [Ne~\textsc{v}] 
(14.3\mum) and [O~\textsc{iv}] (25.9\mum), are not detected 
(only upper limits are reported also in [15]); moreover, the [Ne~\textsc{iii}]/[Ne~\textsc{ii}] line 
ratio of $\sim$0.2 is well consistent with a SB-dominated radiation field. 
As a result, taking into account only mid-IR emission lines would lead to a misclassification of \ir20551\ as a 
pure SB source. The lack of high-ionization 
lines and low [Ne~\textsc{iii}]/[Ne~\textsc{ii}] ratio can be actually reconciled with the presence of a deeply 
obscured AGN by allowing for a peculiar geometry of the gaseous/dusty absorber. Indeed, a large covering factor 
of the putative torus predicted by AGN unification models (Antonucci 1993 [22]) can even prevent the formation 
of the narrow-line region and the production of high-ionization species. The geometrical properties of the 
absorber in a ULIRG is likely much more complicated, and a cocoon-like structure can be reasonably expected. 
Also the other standard diagnostic ratio [S~\textsc{iii}]($\lambda$18.71)/[S~\textsc{ii}]($\lambda$33.48) has 
an intermediate value among ULIRGs ($\sim$0.7), and tends to confirm the latter interpretation.\\
Four lines from pure rotational transitions of warm \h2\ are clearly detected (see Fig.~\ref{fig2}, Tab.~1 
and [15]): 
0-0S(3) 9.67\mum, 0-0S(2) 12.28\mum, 0-0S(1) 17.04\mum\ and 0-0S(0) 28.22\mum. 
The upper levels of these transitions are populated via UV pumping, formation of \h2\ in excited states or collisional 
excitation; therefore these lines directly probe the warm component of the molecular gas. 
A standard ortho-to-para ratio of 3 is found for gas with typical temperature T$\sim 300$~K. 
The heating mechanisms can be associated to either the SB [e.g. in photo-dissociation regions (PDRs), 
shocks/outflows in supernova remnants (SNRs)] or the AGN (due to the X-ray heating). From the line ratios and 
excitation temperatures measured among ULIRGs (including \ir20551) Higdon et al. 2006 [23] ascribed the warm \h2\ 
component to PDRs associated with massive SB. A more detailed investigation of the physical parameters of the 
\h2\ gas is presented in Section~3.2.
\section{The circumnuclear medium}
The combined analysis of 3--8\mum\ data gives the immediate advantage to trace the co-existing AGN and SB 
environments. Indeed, after the review of all the mid-IR spectral properties the presence of a heavily absorbed 
AGN combined with a vigorous SB in \ir20551\ is well established. None the less, a comprehensive interpretation 
of all the observables (AGN 
hot-dust emission, continuum reddening, absorption features, PAH strength) is not straightforward. The general 
picture is complicated by the different spatial extent of the nuclear region that has been explored in the works 
mentioned above. In fact, there can be some aperture effects related to the slit widths, as the nuclear emission 
is quite diffuse and has a large surface brightness. The slit widths and orientations of the main instruments 
considered in this work are shown in the right panel of Fig.~\ref{fig0}.
\ir20551\ presents a very small fraction ($< 10$\%) of extended emission in the 13.2-$\mu$m continuum, which 
can be mainly associated with the compact, unresolved hot/warm dust component in proximity of the AGN 
(D{\'{\i}}az-Santos et al. 2010 [24]). Conversely, the extra-nuclear emission is substantial for both the 
7.7-$\mu$m PAH feature and the [Ne~\textsc{ii}] line at 12.8\mum\ ($\sim$40 and 25\% respectively; 
D{\'{\i}}az-Santos et al. 2011 [25]), which are obviously related to the circumnuclear SB. 
Here, in order to further investigate the physical conditions responsible for reddening/absorption we try  
(\textit{i}) to fit simultaneously the \textit{L}-band and 5--8\mum\ data, and (\textit{ii}) to measure the 
column density of the circumnuclear gas for both the atomic and molecular components. 
\subsection{The extinction law}
\begin{figure}[!h]
  \centering
   \includegraphics[angle=0,scale=0.7]{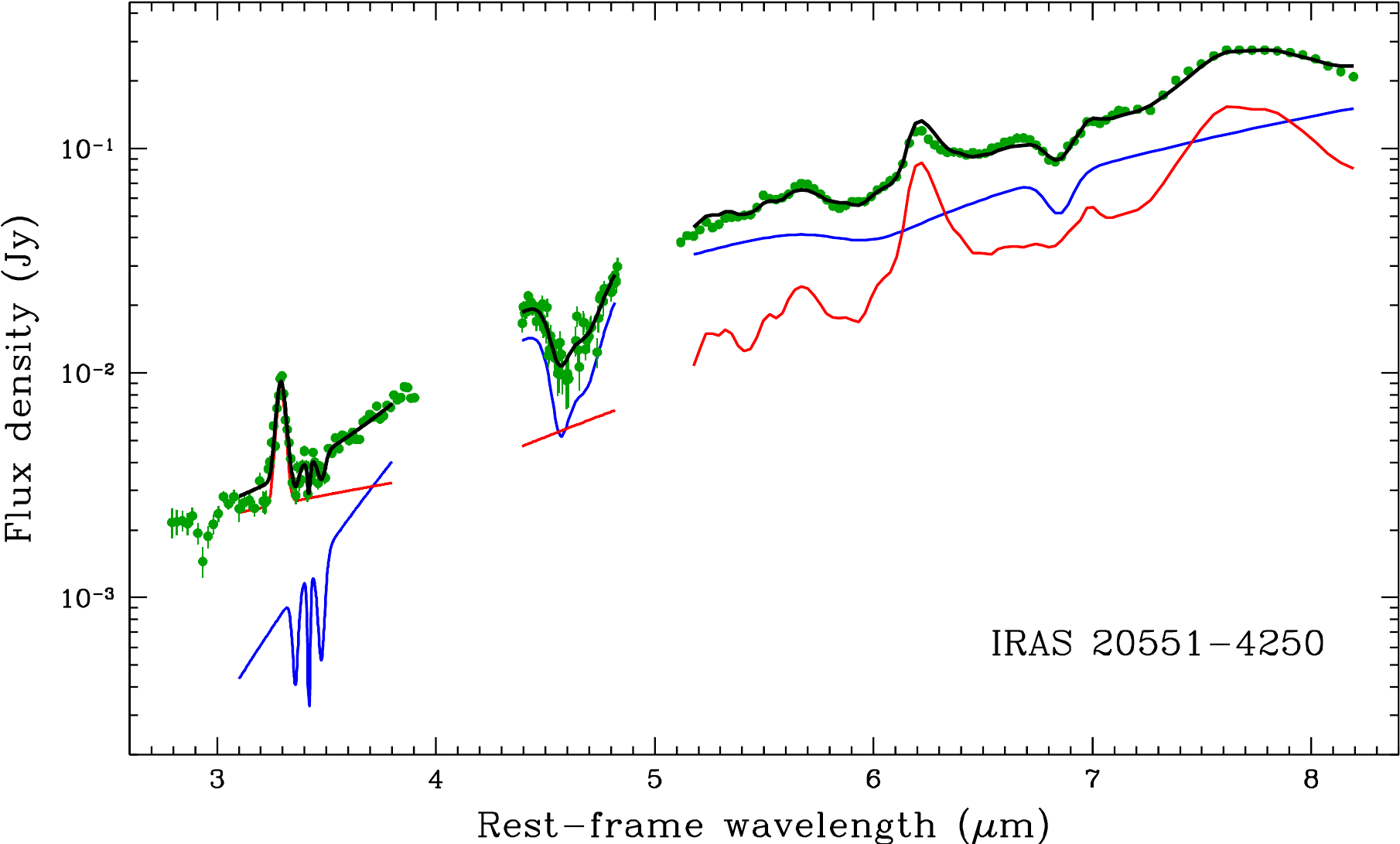}
    \caption{\footnotesize{\textit{VLT}/ISAAC \textit{L}- and \textit{M}- bands, plus \textit{Spitzer}/IRS low 
resolution spectra of \ir20551. The data and best-fit models for each band are plotted in green and black, 
respectively. The different contributions to the observed spectra (blue for AGN component, red for the 
SB one) are also shown. Absorption features are fitted separately and then combined with the AGN template. 
Due to the strong reddening of the AGN continuum, the only AGN signature within the \textit{L}-band is the 
shallow, structured 3.4\mum\ hydrocarbon absorption, highly diluted by the SB emission. The \textit{M}-band 
continuum cannot be constrained due to the deep CO features, while the 6.0\mum\ water ice and 6.85\mum\ 
hydrocarbon absorption profiles are detected at 5--8\mum\ [9,15].}}
\label{fig3}
\end{figure}

\begin{figure}[!h]
  \centering
   \includegraphics[angle=0,scale=0.7]{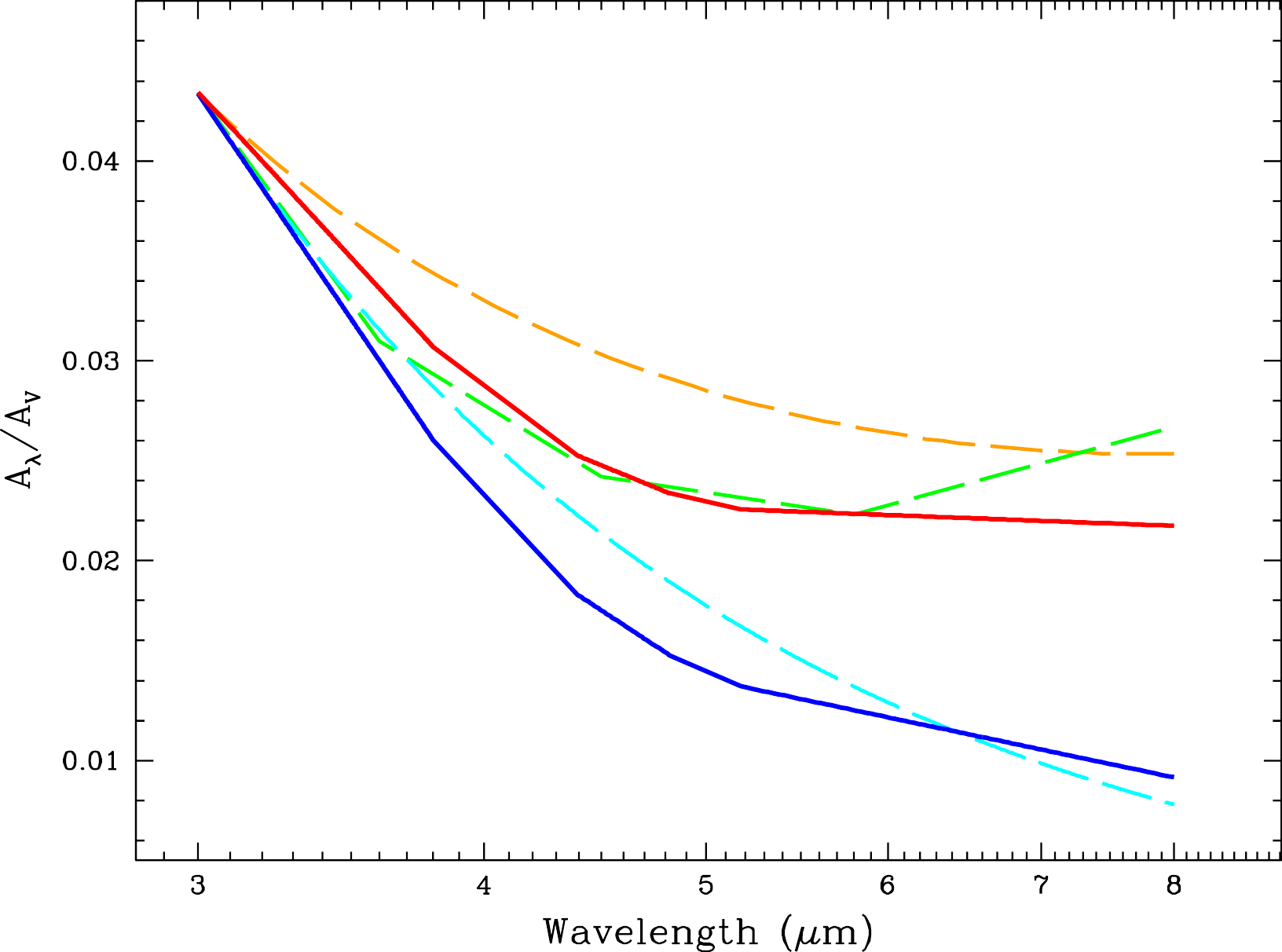}
    \caption{{\footnotesize Dust extinction laws obtained by assuming a single power-law form 
(f$_\nu \propto \lambda^{\Gamma}$) to reproduce the intrinsic AGN hot-dust emission over the 3--8~$\mu$m band. 
The solid lines correspond to different choices of the spectral index: blue for $\Gamma=2$, red for $\Gamma=3$. 
The dashed lines are standard Galactic extinction curves for comparison: cyan for [18], orange for 
Chiar \& Tielens 2006 [28], green for Nishiyama et al. 2008, 2009 [29,30]. 
The relation with the extinction in visual magnitudes plotted on the vertical axis is based on the latter 
works, and all the curves are normalized in order to have the same value at 3~$\mu$m. While the exact extinction 
shape is not so important in the \textit{L}-band, at 5--8\mum\ (and most likely beyond, with the presence of the 
silicate absorption feature) the difference in terms of optical depth in the different cases can be as large as 
a factor of 3. However, it seems quite hard to obtain such a large dust extinction around 6\mum\ to be 
consistent with the gas column density of nearly $\sim 10^{24}$ cm$^{-2}$ measured in the X-rays.}}
\label{fig4}
\end{figure}

Fig.~\ref{fig3} shows the observed spectrum of \ir20551\ between 3 and 8\mum\ once the ground-based 
\textit{VLT} data are combined with the first part of the Short-Low \textit{Spitzer}/IRS orders. We did not 
apply any cross scaling factor, since it would be a very complex task and we are confident about the reliability 
of the absolute flux calibrations, which are affected only by small relative errors ($\sim$10\%; [7,11,17]). 
From a visual inspection of the three spectra it is clear that the observed continuum slope, which is 
expected to be heavily shaped by the AGN contribution, cannot be reproduced with a single spectral index 
over the whole range under investigation. In our separate \textit{L}-band and 5--8\mum\ studies we have assumed 
an intrinsic slope of $\Gamma=1.5$ for the AGN hot-dust continuum, and then applied a power-law extinction of 
the form $\tau(\lambda) \propto \lambda^{-1.75}$ [18]. This screen-like absorption is 
possibly due to colder dust in the outer layers of the putative torus, or it might be associated with 
some star-formation region in the circumnuclear environment of the host galaxy. It is now evident that 
the latter assumptions do not allow us to reproduce simultaneously the AGN emission for the different 
data-sets. In fact, by extending the best-fitting AGN model from the \textit{L}-band to longer wavelengths we 
largely overestimate the 8-$\mu$m observed flux. Of course, it is possible that the intrinsic 
AGN spectrum is more complex than the one adopted in our spectral decomposition. A more detailed 
analysis should allow for different dust components with individual temperature and emissivity, and 
also radiative transfer effects need to be taken into account. However, a broken power-law trend 
seems to describe with fairly good precision the observed spectral curvature. Interestingly, we can 
try to obtain some empirical (\textit{a posteriori}) indication about the extinction suffered by the AGN 
hot-dust emission. Virtually all the available extinction curves in this wavelength range, in fact, are derived 
from lines of sight within our own Galaxy, while the composition of the interstellar medium (ISM) in active 
galaxies is expected to be very different, as proved e.g. by the dust-to-gas ratios estimated through 
a comparison between the mid-IR dust obscuration and the gas column density in the X-rays of these 
objects (Maiolino et al. 2001 [26]; Nardini \& Risaliti 2011 [27]). \\
We have therefore fitted all the three bands allowing for different slopes of the observed AGN continuum. 
The \textit{M}-band is clearly 
poorly constrained and the value of $\Gamma$ is frozen to give a smooth connection among the spectral intervals 
for both the AGN and SB templates. We have then computed the trend of the extinction law by making 
the easiest assumption about the intrinsic shape of the hot-dust emission, i.e. the simple 
power-law dependence of the flux density from wavelength. Fig. \ref{fig4} shows the comparison 
between two possible extinction laws, 
corresponding to different values of the intrinsic $\Gamma$, and three standard Galactic curves. 
Although no conclusive indication can be drawn, the similarity is quite remarkable, and suggests 
that the dust extinction law and the AGN intrinsic continuum are partially degenerate. This anyway 
does not affect the quantitative results of our analysis, as the AGN and SB 6\mum\ to bolometric corrections 
are averaged over large samples and this systematic effect is greatly reduced (see also the discussion on 
the AGN template and dust extinction in [11]).

\begin{table*}
 \begin{center}
 \begin{footnotesize}
 \begin{tabular}{lccccc}
 \multicolumn{6}{c}{Extinction estimates}\\
 \hline
  Band    & $\tau_L$ & $\tau_6$ & $\tau_{3.4}$ &  N$_H$     & expected A$_V$   \\
              &                &                &                   &   cm$^{-2}$  & mag                      \\
  \hline
  3\mum\ [9] & 8    &           &                    &                    & 220         \\
  6\mum\ [15] &           & 1.2     &                    &                   &  110      \\
 3.4\mum\ [9]  &         &           &      1.5          &                   & 450 \\
 2-10 keV [11]  &          &           &                    & $8\times10^{23}$ & 420\\
  \hline
  \end{tabular}
   \label{tab2}
  \caption{\footnotesize{Extinction obtained by assuming a gas-to-dust ratio as in Eq.~\ref{eq1}. 
  Columns: (1) observational band used for the direct measurement with the relative reference. 
  (2)-(3) \textit{L}-band and 6\mum\ optical depth for the AGN continuum as obtan with our spectral 
decomposition model. 
  (4) Optical depth of the aliphatyc hydrocarbon feature once corrected for the SB dilution. 
  (5) Gas column density measured from the X-ray spectrum.
  (6) Optical extinction predicted by Galactic relations (see text for details).}}
    \end{footnotesize}
 \end{center}
\end{table*}
\subsection{Gas and dust content}
To further constrain the absorbing/emitting medium in \ir20551, we attempted at estimating the gas column density 
by means of a multi-wavelength approach. We start by assuming a Galactic gas-to-dust ratio [31], 
\begin{equation}
\frac{N_H}{A_V}=1.9\times10^{21} mag^{-1} cm^{-2}
\label{eq1} 
\end{equation} 
with $A_L\sim0.04~A_V$ and $A_6\sim0.012~A_V$. We then employ the following estimates: 
\textit{(a)} the column density of the gas absorbing the X-ray radiation directly measured in the 2--10~keV 
energy range [14]; \textit{(b)} the \textit{L}-band and 6\mum\ optical depth assessed through the continuum 
reddening in our decomposition method [11]; \textit{(c)} the optical depth of the 3.4\mum\ hydrocarbon 
feature [7]. The corresponding visual extinction values are listed in Tab.~2. \\
From a comparison among these independent $A_V$ predictions, we can draw four main considerations. 
\textit{(1)} Independently from the adopted proxy, we infer a huge extinction in the visual band, which 
naturally explains the optical misclassification of \ir20551. 
\textit{(2)} As discussed in the previous section, a flatter extinction law over the 3--8\mum\ range with 
respect to a steep power-law trend [18] seems to be more appropriate to reproduce the observed AGN emission. 
Otherwise, the values of $A_V$ derived from the 3-$\mu$m and 6-$\mu$m reddening differ by a factor of two. 
\textit{(3)} By using the depth of the hydrocarbon feature to de-absorb the continuum, following the 
Galactic relation $A_L=(12\pm4)\tau_{3.4}$ (Pendleton et al. 1994 [32]), the resulting AGN intrinsic luminosity 
would exceed the source bolometric emission. The abundance of hydrocarbons dust grains is therefore higher in 
\ir20551\ than in the Galactic ISM. 
\textit{(4)} The X-ray column density corresponds to an $A_V(X)$ at least a factor of two larger than that 
expected from our mid-IR modelling ($\tau_L, \tau_6$). Irrespectively of the actual dust extinction law, any 
reasonable value of the mid-IR optical depth implies a lower dust-to-gas ratio than in the Milky Way ISM. 
As a ULIRG is by definition a dust-rich system, this apparent inconsistency can be explained in two ways, which 
are in part complementary: (\emph{i}) due to orientation effects, our line of sight pierces through the regions 
of highest column density in the circumnuclear absorber. (\emph{ii}) There is little coupling between the dust 
and gas components because the bulk of X-ray absorption occurs close to the central engine, in a region 
comparable in size with the dust sublimation radius.\\
\begin{figure}[!h]
  \centering
   \includegraphics[angle=0,scale=0.7]{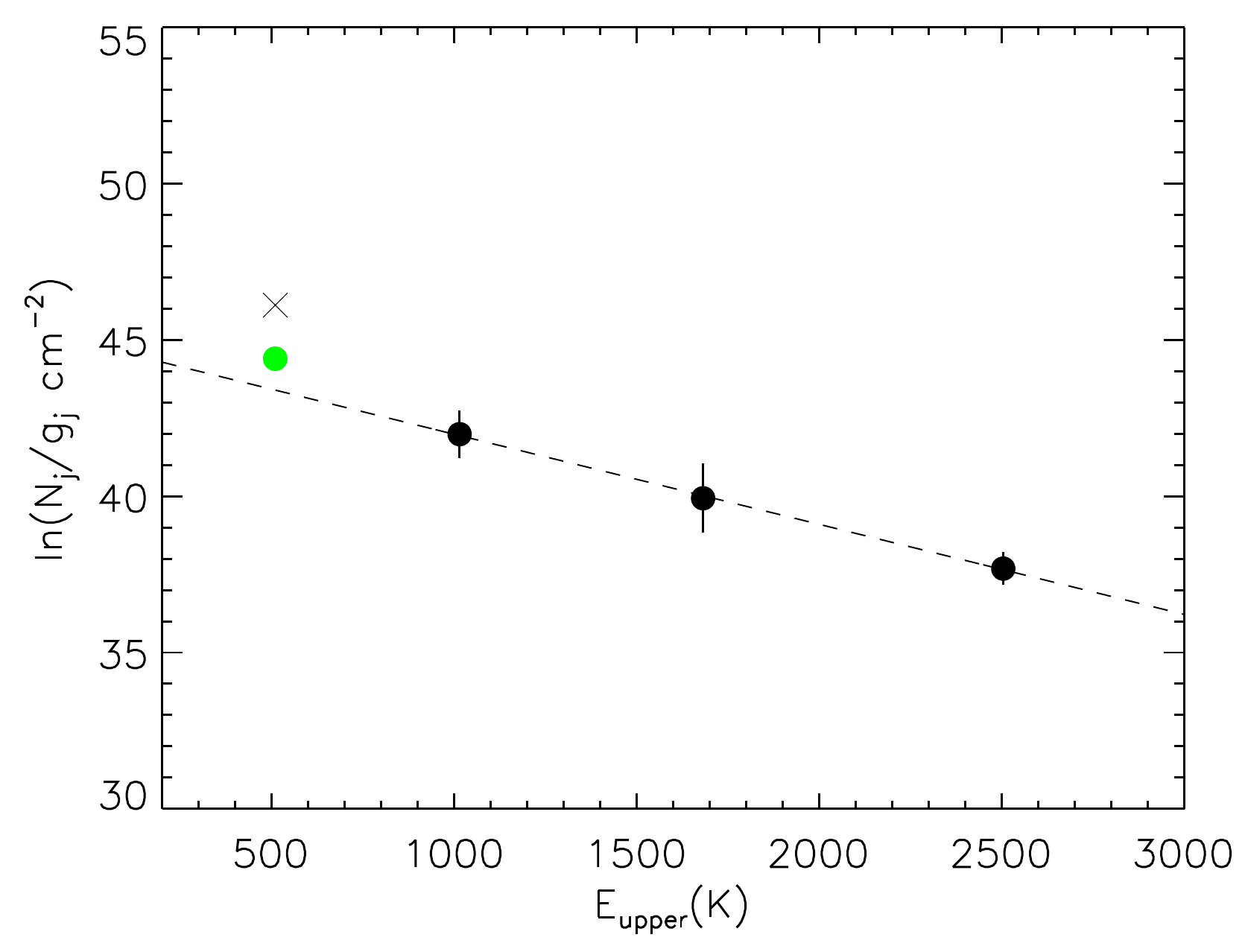}
    \caption{{\footnotesize H$_2$ rotation diagram: for each transition the column density N$_j$ normalized to 
    the statistical weight of the state ($g_j$) is plotted as a function of the upper level energy (Nisini et al. 2010 [34]). 
    The dashed line represent the best LTE fit obtained for S(1), S(2), and S(3) transitions 
    (black points). 
    For completeness we plot also the S(0) observation (black cross) and the measurement corrected for aperture effects (green point).}}
\label{fig5}
\end{figure}
Another line of investigation into the physical properties of the circumnuclear medium relies on the rotation 
diagram of the warm molecular hydrogen, from which we can derive the temperature, column density and mass of 
the gas. To this purpose, the observed fluxes listed in Tab.~1 are converted into column densities of the 
$J$th state ($N_j$) assuming the LTE regime, an ortho-to-para ratio of 3, a point-like source, and no extinction 
[23] (see also Veilleux et al. 2009 [33]). While Higdon and collaborators [23] construct the rotation diagram 
for \ir20551\ using only the \h2\ S(1) and S(3) transitions detected in the low-resolution mode, here we make 
use of high-resolution detections and add the S(0) and S(2) lines. In this way, the parameters derived from the 
linear fitting in Fig.~\ref{fig5} are more reliable and accurate. Clearly a single temperature model applies to 
the S(1), S(2) and S(3) transitions, with the excitation temperature ($T_{\rm{ex}}$) given by minus the 
reciprocal of the slope, while the total \h2\ column density (N$_{\rm H_2}$) depends from the fit 
normalization and the partition functions of the populations. We thus obtain 
$T_{\rm{ex}}=347^{+5}_{-6}$~K, $N_{\rm H_2}=(2.7\times10^{20})$~cm$^{-2}$ 
and a corresponding \h2\ mass of $M_{\rm H_2}=6.8\times10^8 \rm M_\odot$.\footnote{Uncertainties on 
$T_{\rm{ex}}$ and $N_{\rm H_2}$ 
are estimated by re-fitting the data while varying the S(1) and S(3) fluxes within their errors.} 
Our estimate gives a higher temperature ($\sim 8\%$) and correspondingly lower gas mass with respect to [23]. 
The inclusion of the S(0) line require some caution, as it is detected with the IRS-LH slit, 
much larger ($11.1\arcsec$) than the SH one ($4.7\arcsec$) that samples the previous fluxes. 
For completeness, we plot in Fig.~\ref{fig5} the observed S(0) value as a cross and the value corrected for the 
relative slit apertures SH/LH as a green point. Including also the corrected S(0) significantly steepens the 
linear regression and leads to a lower temperature T$_{ex}=303$~K, hence doubling the column density and mass. 
As a matter of fact, a single-temperature component is not suitable to properly reproduce complex systems such 
as \ir20551, and a multi-temperature model should be adopted [23,33]. Unfortunately the non-detection of higher 
level transitions [e.g. from S(4) to S(7)], or their blending with PAH features, prevents us from modelling a 
hot ($T \simeq 1000$~K) \h2\ component. None the less, as an exercise, we can exclude the S(3) point and adopt 
the corrected S(0) in the linear regression. We now trace a colder \h2\ component with $T_{\rm{ex}}=265$~K, 
characterized by a huge, likely unphysical\footnote{A mass for the molecular hydrogen larger 
than $\sim 10^9 \rm M_\odot$ would correspond to enormous star formation rates, with an IR radiation even 
greater than \ir20551\ bolometric luminosity.} gas mass ($M_{\rm H_2}\sim 2\times10^9 \rm M_\odot$). 
We remind that ortho-\h2\ exists only in states of odd rotational quantum number, while para-\h2\ is represented 
only by states of even $J$, therefore the S(1)/S(3) line ratio is independent from the ortho-to-para ratio. 
The measured S(1)/S(3) is $1.29\pm0.07$, in agreement with the theoretical value of 1.23 computed for 
\textit{no} extinction and $T_{\rm{ex}}=350$~K. 
From this, we conclude that the obscuring material along the line of sight producing the continuum reddening, 
deep features and X-ray absorption lies in between the AGN and molecular \h2\ clouds and is possibly associated 
with the SB region. 
\begin{figure}[!h]
  \centering
   \includegraphics[angle=0,scale=0.5]{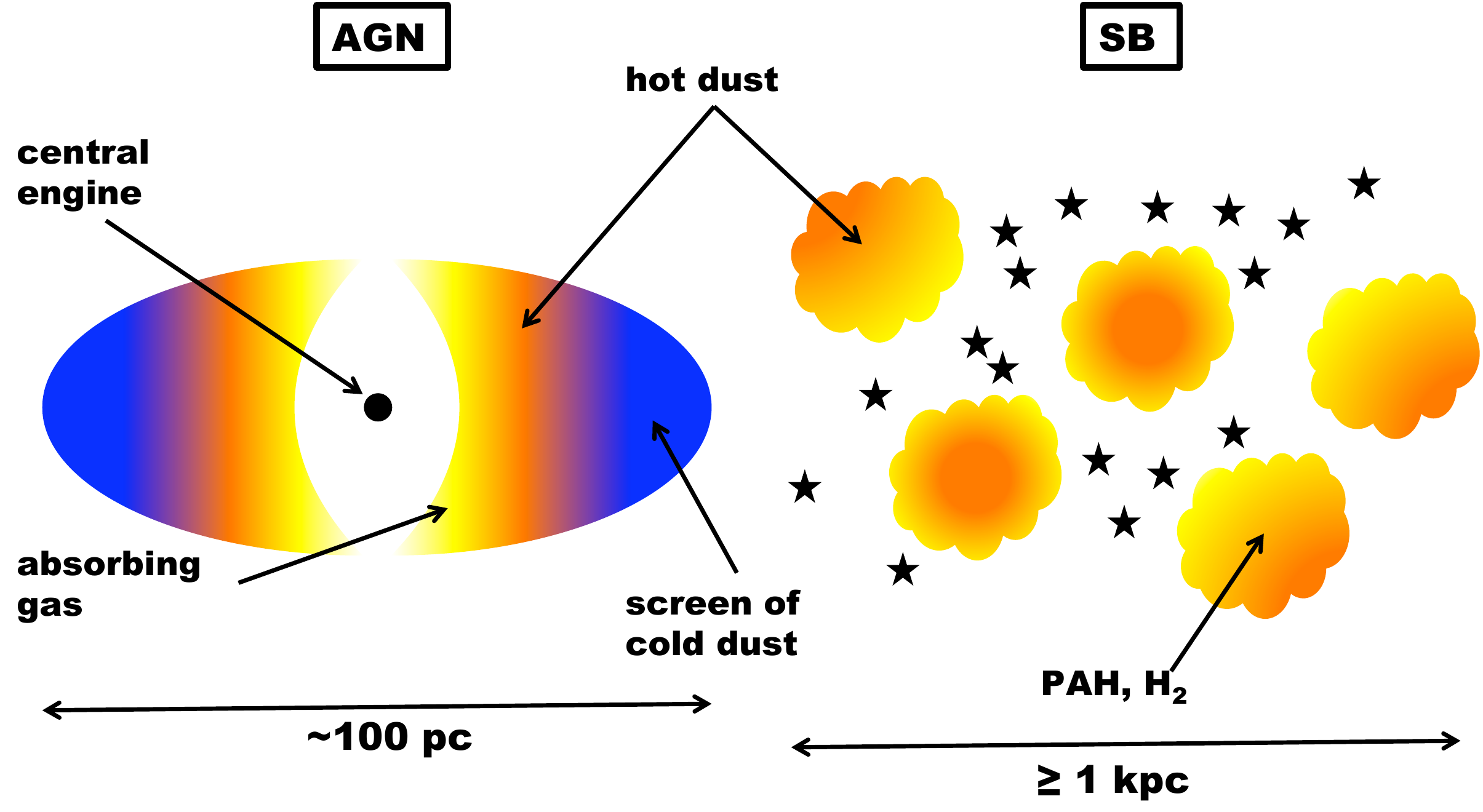}
    \caption{{\footnotesize Schematic view of the possible spatial distribution of the absorbing/emitting medium in \ir20551. 
    The X-ray absorbing gas (in light yellow) is located within the dust sublimation radius and represents the inner edge of the axysimmetric (toroidal) absorber. 
    The dusty regions emitting the thermal radiation are in orange (for both AGN torus and SB clouds). 
    A compact, dense screen of cold dusty gas (in blue) forms the outer layer of the AGN absorber, and is responsible for both continuum reddening 
    and deep absorption troughs. PAH features and \h2\ lines come from diffuse material spatially mixed with young stars (yellow/orange clouds). 
    Due to the large spatial extent, only moderate internal obscuration is affecting the SB environment. }}
\label{fig6}
\end{figure}
\section{Discussion}
We can now compile all the different aspects of the previous analysis in order to construct a comprehensive 
picture of the absorbing/emitting medium in \ir20551. A stratified structure of the circumnuclear material, 
involving the different spatial scales (see Fig.~\ref{fig0}), can well explain all the observational 
evidence. The basic ingredients are summarized as follows: \textit{(i)} the hot dust component, where the 
grains are transiently heated to temperatures close to the sublimation limit, can be associated 
with both the inner surface of the AGN torus and the starburst environment; however, due to the different 
spatial concentration of the hot dust in the two cases, the resulting nearly power-law continuum is much more 
intense for an AGN. \textit{(ii)} A cold dust component and a large amount of gas are required to produce the 
continuum reddening, the deep absorption features (aliphatic hydrocarbon, CO, HAC, silicates), as well as the 
X-ray absorption. Consequently, the inferred properties of the circumnuclear absorber point to an optically 
thick screen along the line of sight towards a point-like source such as a bright AGN, rather than to a diffuse 
dust distribution spatially mixed with the energy source (as in a starburst). Moreover, this dust screen must 
be geometrically thick, since a large covering factor would be consistent with the absence of high-ionization 
coronal lines (e.g. [Ne~\textsc{v}], [O~\textsc{iv}]). These properties are typical of the AGN putative torus, 
which is located at the spatial scales ranging from a few pc to several tens of pc from the central engine. 
The obscuring medium is also expected to be sufficiently close to the central AGN (i.e. with the inner edge of 
the torus falling within the dust sublimation radius) to allow for the observed gas over-abundance. 
On farther scales (several hundreds to a few thousands pc), molecular clouds are associated with the starburst 
event. Here, in addition to warm thermal dust, the PAH grains can survive and give rise to the typical set of 
emission features usually employed as SF tracers. Furthermore, with the increasing optical depth within the 
individual star-forming clouds, photo-dissociation becomes eventually slow and inefficient, so that hydrogen 
also appears in the molecular state. This explains the unextincted \h2\ pure rotational lines detectd in the 
mid-IR. A cartoon of the circmunuclear environment is shown in Fig.~\ref{fig6}. \\
Of course, the qualitative considerations driven by the mid-IR spectral properties are not sufficient to 
fully understand the multiple physical conditions characterizing such an extreme source. In order to probe 
the nuclear enviroment and its surroundings, a detailed spectral analysis at different wavelengths is needed, 
possibly resolving and disentangling the different spatial scale. This would make it possible to address the 
problems connected to the uncertain shape of both the intrinsic and the observed AGN continuum, and therefore 
to better constrain the actual extinction law. At present, even the joint modelling of the $\sim$2--20\mum\ 
spectral energy distribution (SED) is frustrated by the spread of the signal-to-noise ratio (S/N )and the 
relative flux calibration among ground-based and space facilities involved in the observations. The forthcoming 
\textit{James Webb Space Telescope} (\textit{JWST}) is the ideal instrument to probe the mid-IR SED of local 
ULIRGs, offering the opportunity of high-quality data obtained with relatively short exposures. For example, 
a high-resolution ($R \sim 2700$) observation of \ir20551\ with NIRspec (Bagnasco et al. 2007 [35]) centred 
at 3.5\mum\ requires only 
$\sim$300 sec of exposure time\footnote{We used the exposure time calculator available at 
http://jwstetc.stsci.edu/etc/input/nirspec/spectroscopic/ with the following settings: G395H grating plus 
F290LP filter, average thermal background and zodiacal light.} to reach a S/N$\sim 150$ per resolution element. 
At longer wavelengths, the medium-resolution spectrometer MIRI (Wright et al. 2004 [36]) will ensure similarly 
high performances. 
Besides the unique settings available (among which integral field unit and multi-shutter array), \textit{JWST} 
will fully cover the $\sim$1--25\mum\ range, allowing us to detect and resolve even faint and/or blended 
features. In this context, the separation of highly excited rotational levels of the CO $\nu=1-0$ band would be 
particularly suitable to constrain the dense gas temperature, density and kinematics within the circumnuclear 
environment (see e.g. Shirahata et al. [37]). 
\section{Conclusions and remarks}
In the present work we have first reviewed the properties of \ir20551\, a prototypical local ULIRG observed by 
our group in the \textit{L}- and \textit{M}-band with ISAAC at the \textit{VLT}. The spectral analysis also 
includes the 5--8\mum\ spectrum obtained by \textit{Spitzer}/IRS. According to the AGN/SB decomposition method 
we have developed in several previous papers [7,10,17], \ir20551\ turns out to be a composite source, dominated 
in the mid-IR by hot dust emission associated with deeply embedded BH accretion and characterized by a 
vigorous circumnuclear starburst which provides the main power supply to the whole system. We have then 
interpreted the key spectral properties of the source over the $\sim$3--20\mum\ wavelength range (e.g. the 
reddening of the continuum, the presence of deep absorption features, the lack of high-ionization coronal lines 
and the detection of \h2\ rotational transitions) in the framework of dust and gas spatial distribution and 
physical conditions. Our main results are the following: (\textit{i}) the shape of the AGN intrinsic continuum 
is partly degenerate with the form of the extinction law. This is mainly evident beyond 5\mum. 
(\textit{ii}) Given the gas amount inferred from X-ray observations, the central regions of \ir20551 seem to have 
a dust-to-gas ratio much lower than the Galactic interstellar medium. (\textit{iii}) Aliphatic hydrocarbon and 
HAC grains are over-abundant with respect to the local molecular clouds. (\textit{iv}) A large covering of the 
nuclear engine likely prevents the ionization of the AGN narrow-line region and the excitation of fine-structure 
lines. Therefore, a screen of cold, dusty gas lies along the line of sight to the AGN, heavily extinguishing its 
spatially compact primary emission. (\textit{v}) A large amount ($\rm M_{H_2}=6.8\times 10^8 \rm M_\odot$) of 
warm ($T_{\rm{ex}}=347$~K) molecular hydrogen and PAH grains are associated with the starburst environment on 
typical scales of a few kpc. The findings have been \textit{qualitatively} interpreted by means of a simple 
geometrical configuration as the one sketched in Fig.~\ref{fig6}. We have finally described the great 
improvement in terms of sensitivity, spectral coverage and resolution that will be achieved in the near future 
with the advent of \textit{JWST}. This will also allow us to separate the different spatial scales and explore 
in larger detail the connection between the AGN and SB environments and the mutual feedback between the two 
physical processes.

\section*{Acknowledgements}
We thank the anonymous referee for the constructive comments and suggestions. \\
E. Sani is grateful to Dr. F. Fontani for precious discussions on the star forming environment. 
This work has made use of the NASA/IPAC extagalactic database  (NED). 
ES acknowledges financial support from ASI under grant I/009/10/0/. 
EN acknowledges financial support from NASA grants NNX11AG99G and GO0-11017X. 
\newpage
%


\end{document}